Research Article

# HydroPower Plant Planning for Resilience Improvement of Power Systems using Fuzzy-Neural based Genetic Algorithm

Akbal Rain [1], Mert Emre Saritac [2*]

[1] *Department of Electrical and Computer Engineering, Southern Illinois University of Carbondale, USA*
[2] *Department of Computer Science, University of Massachusetts Dartmouth, USA*

## 1. Introduction

Environmental concerns due to the emissions of natural gas power plants as well as high investment costs of these assets render power systems to be operated mainly by renewable energy resources such photovoltaic, wind, hydropower, etc. [1]. At recent decade, small hydro-power used in developing countries which plays a vital role in rural electrification. This microgrid is one of the renewable energies which has some advantages which listed below [2]:

I) Storage time is low due to lack of heat

II) There is no pollution when this energy working

III) Low cost with more reliability and reducing transmission losses due to no long transmission network

IV) There is no need to stand connected loads; instead, there is a growing gap from supply to monthly electricity demand.

Small-hydropower has the best potential in the most areas of the world which consider low cost with more benefit for new green energy for futures [3]. Small-hydropower has faced with new technology in 2015 to 2019 in control part which will be review at the continue of this article, but it also considers some challenges for more stability, load frequency control (LFC), and new kinds of intelligent controllers [4, 5].

Eliminating the mismatch between production and load keeps the system frequency constant.

The most important aspect of operation of electric power systems is to be resilient and controllable in the case of disturbances [6]. These controllers need to be reliable; in other words, they must manage load fluctuations and system disturbances in limitation of voltage and frequency [7]. In this study, a new model proposed for more stability of hydropower systems in small scale which use Secondary Load Controller (SLC) for dividing extra power generated by the power plant. This happened when frequency is sensing in the small-hydropower. The SLC always have loads in generator and also in turbine side, due to small-hydropower has some low in terms of working. The main goal of this study is to provide simple operation at the system level with low cost and low maintenance with more reliability and stability and also plug-and-play mode to install it at the any parts of electrical systems. So, using SLC can solve many problems in small scale hydropower system about controlling frequency and voltage in any area with minimum cost and the best performance. The controller of small-hydropower system is PD which is varied at any times. This state of PD controller, created an uncertainty which require to use fuzzy logic to tuning PD controller. Fuzzy logic used due to presenting membership function and accurate measuring. At the other side, SLC needs a quick response to adjust the frequency during high current periods in order to turn off the diesel production system. So, Genetic Algorithm optimize fuzzy PD controller.

Because of needing robustness and more adaptive controller, the essential use of Fuzzy logic is obvious which can be provide more flexibility in decision-making process to interact between machines-to-machines and machines-to-humans. Also a new model proposed which can optimize Fuzzy-PD controller in LFC terms based Genetic Algorithm and then neural deep learning technique which used Deep Spiking Neural Network (DSNN) which can be provide more robustness, stability and reliability and also low cost.

## 2. Literature Review

In [8] a new analysis performed in the process of a sudden increase of load for a hydropower plant. The main focus of this study is to investigate the differences between stable vibration and the instability of the hydropower system for each region to estimate the error tolerance. In [9], a nonlinear load frequency controller by using State-Dependent Riccati Equation (SDRE) methodology proposed for hydropower plants. The main goal of this article was obtaining a desire frequency in any condition such as nominal and uncertain mode. In the simulation, a comparison between proposed schema and conventional controller presented which Investigating the effects of noise and uncertain parameters in the systems.

There are many approaches which studied about hydropower. For example, in [10], for the first time an experiment model of hydropower for frequency stability was conducted. According to this article, a cognitive study of Theoretical analysis and physical experiment model combined with each other which results were consistent with the analysis of the theory relative to the experimental mode in a simulation with a tolerance for fault tolerance. In [11], a new approach for load frequency control for small-hydropower proposed which used self-tuning fuzzy proportional-derivative. This microgrid was controlled by secondary load bank due to more stability and gain. Another study which presented in [12], optimized high penetration of voltage in large-scale of hydropower station with new evolutionary algorithm named glowworm swarm optimization. This economic load dispatch method optimized the accuracy and avoided the prematurity in the small scale of hydro-turbine. The simulation results of this study was compared with another namely evolutionary algorithms such as Standard Glowworm Swarm Optimization (GSO), Ant Colony Optimization (ACO), Genetic Algorithm (GA), Particle Swarm Optimization (PSO) and Evolution Programming (EP). The results represented that the proposed method was more robust in large-scale of hydropower station. In [13], load frequency control with the most reduced dump load at the isolated small-hydropower plant proposed. By comparison the results, it presenting that proposed method had better transient performance in comparison to two-pipe case with 50% rate of dump load and also 30% dump load of three-pipe case. In [14], PID controller used for load frequency control at the hydropower systems. PID controller used in this method due to mentioning schemes which was not corresponding between main power system inputs such as change load demand and change in speed turbine settings. This method confirmed under Automatic Generation Control (AGC) in power scheme. More rapid output response and minimal overshoot were the main results of this method.

Pumped hydropower energy storage can be used in renewable energies. This energy storage method was proposed in [15] reference, in integrated microgrid such as hydropower for more robustness and optimization. Due to these, an evolutionary algorithm named Artificial Sheep Algorithm (ASA) was used for load frequency control and optimized energy storage. Also, a new model of load frequency control with hydro turbine in hydropower system was proposed in [16], under various conditions which used PID controller to optimize load frequency control and more gain in any condition and area. In [17], a method based on the ARIMA model is presented for modeling and analyzing electricity demand with the aim of predicting the use of electricity in Pakistan. This article obtained a sensitivity and intelligence analysis of annual GDP and population growth rates and compared future schedule of hydroelectricity generation from 20020 to 2030 which results represented an increasing use of hydroelectricity consumption in 2030 about 23.4%.

In [18], an analysis of reliability and sustainability of hydropower system proposed in Tibetan Plateau in 2016. The results of this study represented that in addition to the maximum efficiency of hydropower systems, it is an essential to minimize the losses of ecosystem services. In [19], a dynamic approach with randomized load analysis is presented in a hydropower storage system. PI controller modeled as nonlinear method with dynamic behaviors and bifurcation diagram of PI in each level of hydropower system drawn to show the stability of system. In [20], a new and optimized controller applied in small hydropower plant which use Particle Swarm Optimization (PSO) with fuzzy sliding mode as controller. In fact, for optimizing fuzzy memebrships, PSO applied to tune fuzzy parts and after developing the controller, Kalman filter used as an estimator to recognize any variable in system to measure parameter values. Another article which presented in [21], used genetic algorithm for optimizing LFC in small hydropower system. The main goal of this study is reducing frequency deviations which result represented 60% Lower than other research. In [22], isolated hydropower system in small scale surveyed and tried to optimize LFC techniques to develop the system in any conditions. In [23], a novel PID controller used with optimal parameters of hydropower system in small scale with regulating quality parameters (RQPs) based on direct solving method (DSM) in no time domain platform in simulation. This system represented better planning of surge tank in hydropower's subsystem. In [24], planning of electricity generation of hydropower systems in Canadian province surveyed and a new model proposed based on energies and costs.

## 3. Problem Definition

### 3.1. A Brief Introducing of Fuzzy Logic

Fuzzy knowledge base system is an expert system which emulates the human expertise in a certain domain based on fuzzy logic instead of Boolean logic. Fuzzy logic architecture contain some scrips such as inputs, Fuzzyfier, reference engine, rules, and Defuzifier [25]. Fuzzy knowledge base system is a kind of fuzzy expert system which use fuzzy sets and rules. The fuzzifier decides how the crisp input will be converted to linguistic variables used by membership functions which stored by fuzzy rules in the knowledge base. The fuzzifier decides how to convert results of the fuzzy inference engine into a crisp value. The inference engine decides how to process the rules in the knowledge base using If-Then type fuzzy rules which these rules can be convert the fuzzy input to the fuzzy output [20]. A fuzzy inference system architecture represents in Figure 1. Fuzzy controller should be robust and reliable, so this article present a novel type-II fuzzy PID controller. This method form a secondary frequency control loop which can be maintain frequency and tie-line power to their nominal values under different uncertainties.

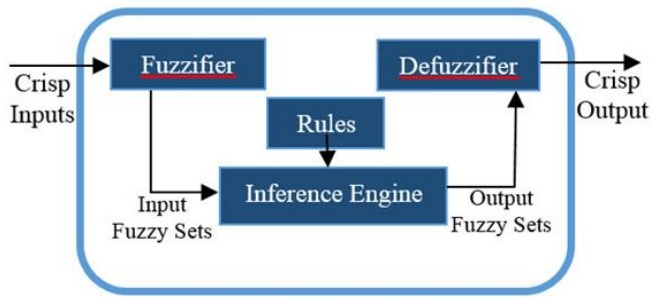

**Figure 1.** Architecture of fuzzy logic system

The small scale of proposed hydropower modeled in MATLAB/Simulink and illustrated in Figure 2. which is the main block diagram of [8] and will be optimize here. The hydropower turbine was modeled as hydropower equation which determined in Eq. (1).

$$P_m = Qhgef \tag{1}$$

In Eq. (1), the parameter P_m is output power of theoretical turbine (kW), ef is the turbine efficiency, Q is flow rate (m$^3$/s), h is the gross head height (m) and g is gravitational constant which fixed in 9.81 m/s$^2$. It is to use synchronous condenser to control the grid voltage at its baseline level. The additional power which collected by Secondary Load Controller (SLC) will be absorbed from small-hydropower system. The Secondary Load Controller (SLC) design which is include an ideal Gate Turn-Off Thyristor (GTO) switches in series mode and also three phase resistive loads in eight sets where these schema varies from 0 to 446.25 kW with a step of 1.75 kW which track binary progression. Also in this schematic, discrete frequency regulator model will be used as frequency control.

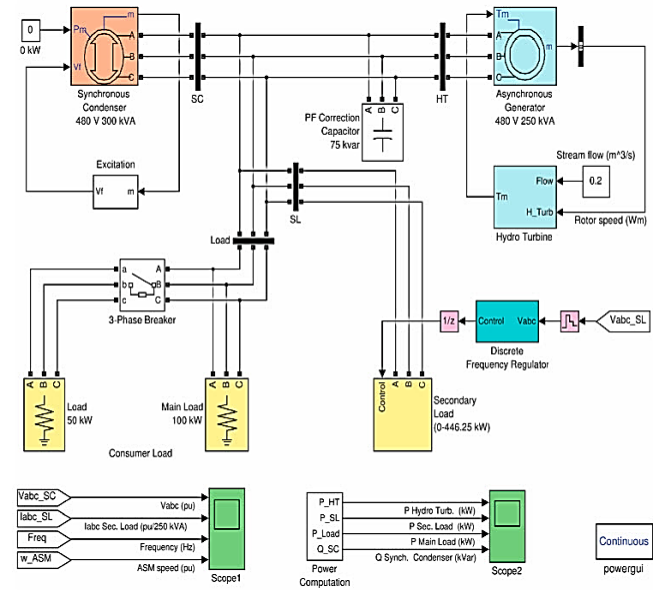

**Figure 2.** Small scale hydropower Simulink model with LFC [11]

### 3.2. Discrete Frequency Control System

The control system of proposed schema is discrete frequency. Secondary load system (SLC) can control the frequency by manipulating. Three Phase Locked Loop (PLL) system can measured frequency and after that in order to estimate frequency error it compared with 50 Hz reference frequency. The secondary load of the small scaled of hydropower system is obtained from the errors summation and then it converted to an eight-bit digital signal to control the three-phase secondary load switches. Zero switching voltage can minimize voltage disturbances in the system by changing it. In this paper, three parts of controllers are presented to optimized the LFC of small scale hydropower system include a conventional PD controller, a new self-tuning fuzzy PD controller and a self-tuning fuzzy PD Geneftic Algorithm controller. These controller performances for setting frequency in dynamic loads will be compared in simulation part.

## 4. Numerical Method

### 4.1. PD Controller Setting

$e(t)$ is an error of PD controller that is obtained from differences between actual output y(t) and desired output *r(t)* to generate a control signal *u(t)* as Eqs. (2) and (3).

$$e(t) = r(t) - y(t) \tag{2}$$
$$u(t) - K_p e + K_D \frac{de}{dt} \tag{3}$$

The PD controller is a controller based on the output feedback. This controller is one of the most commonly used control systems. In the proposed structure, a PD controller along with the operator and evaluator of the proposed model for the hydropower system is used. The PD controller conversion function is given Eq. (4).

$$G_c(s) = K_p + \frac{K_1}{s} + K_D S \tag{4}$$

With respect to Eq. (4), K_p is proportional gain, K_1 is integral gain and K_D is a derivative, which are in fact the same parameters of the PD controller. The physical examination of this controller is possible by an electrical circuit. In order to load frequency control, the PD is used as a compensator along with a fuzzy-based control technique. The reason for using the fuzzy method is because of the uncertainty surrounding load frequency control of hydropower system when the PD controller is used. The derived derivative of the output is represented by Eq. (5) and (6).

$$e(k) = r(k) - y(k) \tag{5}$$

$$u(t) = k_p [e(t) + k_i \int_0^t e(t)dt + k_d \frac{de(t)}{dt}] \tag{6}$$

According to Eq. (5) and (6), $r(k)$ is the reference system input, $y(k)$ is the system output and $e(k)$ is the system error. Also $k_p$, $k_i$ and $k_D$ are PD controller coefficients. To regulate extraction laws of $k_p$, $k_i$ and $k_d$ descending gradient methods are applied to the chain rule to minimize the performance index function. The incremental PD controller algorithm is defined as Eq. (7) to (10).

$$\Delta k_p = -\mu \frac{\partial E}{\partial k_p} = -\mu \frac{\partial E}{\partial y} \cdot \frac{\partial y}{\partial u} \cdot \frac{\partial u}{\partial k_p} = \mu e(k) \cdot \frac{\partial y}{\partial u} \cdot e_p(k) \tag{7}$$

$$\Delta k_i = -\mu \frac{\partial E}{\partial k_i} = -\mu \frac{\partial E}{\partial y} \cdot \frac{\partial y}{\partial u} \cdot \frac{\partial u}{\partial k_i} = \mu e(k) \cdot \frac{\partial y}{\partial u} \cdot e_i(k) \tag{8}$$

$$\Delta k_d = -\mu \frac{\partial E}{\partial k_d} = -\mu \frac{\partial E}{\partial y} \cdot \frac{\partial y}{\partial u} \cdot \frac{\partial u}{\partial k_d} = \mu e(k) \cdot \frac{\partial y}{\partial u} \cdot e_d(k) \tag{9}$$

$$u(k) = u(k-1) + k_p \cdot e_p(k) + k_i \cdot e_i(k) + k_d \cdot e_d(k) \tag{10}$$

In order to uncertainty in the PD controller in load frequency control due to different conditions in the identification of system, it is necessary to use a controller that can guarantee this uncertainty. Therefore, the fuzzy controller is used. Four-phase fuzzy controller has been used to improve the parameters of the PD controller. *j*'th law is given by Eq. (11).

$$Rule\ j:\ If\ x(t)\ is\ N^j\ Then\ u(t) = G_j x(t) \tag{11}$$

According to Eq. (11), $N^j$, the fuzzy period of the law *j* corresponds to the system state vector $x(t), j=1,2,3,4$; $G_j \in R^{1\times 4}$ is the law *j* for the collector feedback vector. The fuzzy controller inference output is given by Eq. (12).

$$u(t) = \sum_{j=1}^{4} m_j(x(t)) G_j x(t) \tag{12}$$

According to Eq. (12), the Eq. (13) and (14), the nonlinear function of the vector *x(t)* and $\mu_N{}^j (x(t))$, *j=1,2,3,4* membership functions that should be designed and fuzzy logic used for reliability optimization.

$$\sum_{j=1}^{4} m_j(x(t)) = 1 \tag{13}$$

$$Fuzzy - PD_{m_j}(x(t)) = \frac{\mu_{Nj}(x(t))}{\sum_{k=1}^{4} \mu_{Nj}(x(t))}\ for\ j = 1,2,3,4 \tag{14}$$

It should be noted that the reliable fuzzy controller does not require $m_j(x(t)) \in [0\ 1]$ for all *j* states.

*4.2. A Brief Introducing of Deep Spiking Neural Network (DSNN)*

Neural networks are usually convert into deep learning models in some condition especially if neural networks have at least two hidden layers of input nonlinear conversion. In this study, only feedback networks are considered that calculate mapping from input to output. Deep Spiking Neural Network (DSNN) were initially studied as biological information processing models in which neurons exchange information through spikes. The main part of DSNN is that all spikes are always assumed as stereotypical events. So as a result, information processing is reduced to two factors include 1) spikes timing, for example movement frequencies, and the relative timing of pre- and post-synapse spikes and also specific movement patterns. Second, the synapses identification used means which nerve cells are connected, whether the synapse is stimulating or inhibitory, synaptic power and short-term plasticity, or modifying effects are possible. Depending on the level of detail of the simulation neurons, both neurons are the point at which input spikes instantly alter their membrane (physical) potentials or model as multi-chamber models with complex spatial (dendritic) structures. So that dendritic currents can interact before that. Physical potential has also been modified. Different models of spike neurons, such as fusion and fire, spiking response or Hodgkin's Huxley model, describe the evolution of membrane potential and spike production at different levels of detail. Typically, the membrane potential of the stream merges with the entry of the spikes, creating a new spike each time the threshold is crossed. After the spike is created, through the axon to all the nerve cells connected with the delay, the small axon is sent and the membrane potential is adjusted to a certain base. Figure 3 shows this.

Direct communication between analog and Deep Spiking Neural Network (DSNN) is established by considering the activation of an analog neuron as equivalent to the rate of firing of a spike neuron, assuming a stable state. Many neuro-metric models have used such rate codes to explain computational processes in the brain. However, spike neural models can also model more complex processes that depend on the relative timing between spikes or the timing of some reference signals, such as network fluctuations. Temporary codes are of great importance in biology, and even a spike or small time-consuming changes in neuron firing may cause different reactions, as most decisions must be calculated before a reliable estimate of the spike. In addition to the biological definition of DSNNs, they have a pragmatic functional representation that in the field of neural engineering, DSNNs are commonly referred to as spikes and are event-based. Here, an event is a collection of digital information that is identified by the origin and destination address of a time marker. Unlike biologically motivated DSNNs, it may have several bits of load information.

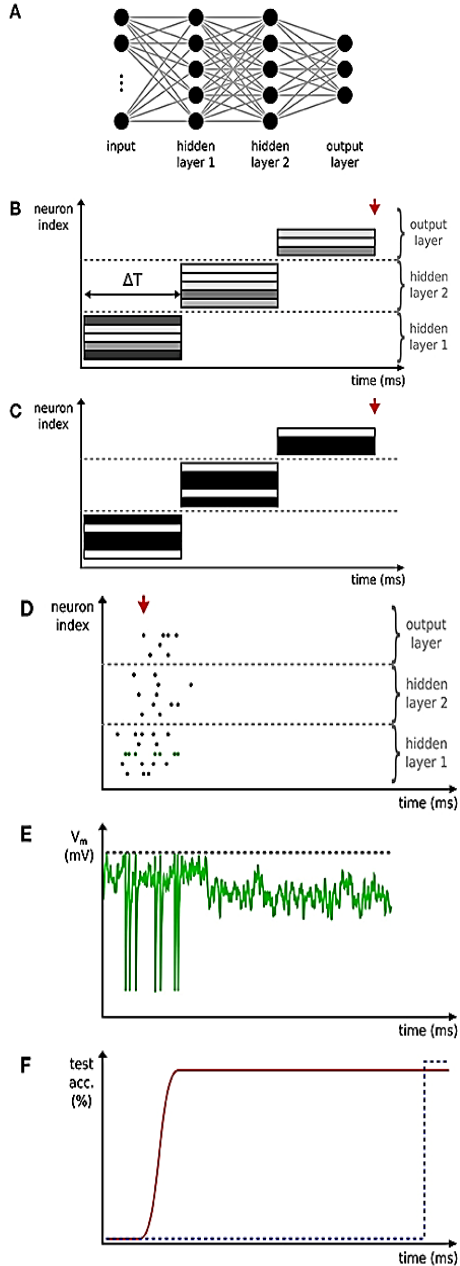

**Figure 3.** General structure and mechanism of Deep Spiking Neural Network (DSNN)

The source of this protocol is the address index or Address Event Representation (AER) protocol, which is used to connect to event-based sensors via digital connection to neural chips or digital hardware after processing. Event-based visual sensors use the loading bit to differentiate between silent and on visual events, but the loading bit can also be used to send other types of information related to post-synapse targets such as Integrate and Fire model [26]. The motivation for studying DSNNs is that the brains show significant cognitive function in real-world tasks. With continued efforts to improve our understanding of brain-like calculations, it is expected that models that are closer to biology will be closer to achieving natural intelligence than more abstract models, or at least more computationally efficient.

In addition to using linear function to activate cells or neurons in the latent layer of nonlinear activator functions such as sigmoid or sinusoidal or functional functions the DSNN presented in this study, due to its high flexibility used non-derivative as well as intermittent activation. By default, DSNN has an equation which presented in Eq. (15).

$$y(p) = \sum_{j=1}^{m} \beta_i \beta_j g(\sum_{i=1}^{n} w_{i,j} x_i + b_j) \qquad (15)$$

According to this equation, $\beta_i$ displays the weights between the input layer and the hidden layer, and $\beta_j$ displays the weights between the output layer and the input layer. $b_j$ it is the value of the neuron threshold in the hidden layer or the bias. $g( ...)$ it is an transition or activation function. The weights of the input layer or $w_{i,j}$ and bias $b_j$, are randomly assigned. The activation function $b_j$ is assigned at the beginning of the number of neurons in the input layer or n and the number of neurons in the hidden layer or m. According to this information, if the known parameters are combined in the general equilibrium and regulated, the output layer will be like Eq. (16).

$$H(w_{i,j}, b_j, x_i) = \begin{bmatrix} g(w_{1,1}x_1 + b_1) & \cdots & g(w_{1,m}x_m + b_m) \\ g(w_{n,1}x_n + b_1) & \cdots & g(w_{n,m}x_m + b_m) \end{bmatrix} \text{ and } y = H\beta \qquad (16)$$

### 4.3. Self-Tuning Fuzzy PD Controller with GA-DSNN

In order to evaluate the efficiency and performance of the PD controller, two fuzzy inputs considered. Two fuzzy inputs are e and ec with some membership function, linguistic variable and fuzzy numbers. linguistic variables are Negative Small (NS), Negative Big (NB), Negative Medium (NM), Zero (ZE), Positive Small (PS), Positive Medium (PM) and Positive Big (PB) based on reference [8]. Table 1 listed the self-tuning modules of the knowledge-based rules. The fuzzy controller rule tables were design based on knowledge of the controller's behavior and through iteration by varying the variables and control parameters. The output of fuzzy controller is u. Now that the fuzzy PD is presented as a controller for the hydropower system, and it is necessary to use the GA-DSNN Algorithm to optimize it. GA-DSNN Algorithm is a combinational and adaptive search technique in evolutionary algorithm-deep learning model that involves some operation such as network generation, network systematic evaluation and network refinement for proposing a solution with potential design until a stopping criterion and finish the networks operations. The GA-DSNN steps execution for optimizing fuzzy-PD controller in hydropower system are as below:

*Step 1*: Read the number of connected modules, insolation model and voltage for each module. In fact, the fuzzy-PD output is GA-DSNN input.

*Step 2*: Define the objective function and recognize the parameters.

*Step 3*: the initial population of chromosome and spikes (chromosomes-neurons) generation.

*Step 4*: Test objective function to evaluate the population

*Step 5*: Convergence examination for testing model runtime. In this part, crossover, mutation and selection operator of GA used. If it satisfied in iteration, then stop continue, and otherwise.

*Step 6*: Start the training/testing process by applying the GA-DSNN such as brightness and in determined iteration to produce next generations.

*Step 7*: Evolving the new generation and go back to step 3.

In this research try to find the optimal membership functions using the GA-DSNN and this is done by the following steps:

Optimization criterion selection: In this work the quadratic criterion will be used to be minimized as Eq. (17).

$$J = \int e^2 \, dt \quad , \quad e = P_{max} - P \tag{17}$$

This selection made for improving the response time and reduce fluctuations. Creation of the initial population: The population consists of a set of individuals, where each individual is composed of the three as listed below:

- Chromosomes and Spikes (Neurons): e, ec and u, each spikes or neurons is composed of a set of genes.
- For Chromosomes and Spikes (Neurons) e the genes are: F1, F2, F3, F4.
- For the ec Chromosomes and Spikes (Neurons) the genes are: F'1, F'2, F'3, F'4.
- For the Chromosomes and Spikes (Neurons) u the genes are: F"1, F"2, F"3, F"4.

A range of variation of F, F ', F " which is the search space is defined which varies between [0.01 0.99]. The resulting of Chromosomes and Spikes (Neurons) production is shown in Table 2. It should be noted that these rules and spike production do not have any nomenclature.

**Table 1.** Knowledge-based rules of fuzzy modules

|    | NB | NM | NS | Z  | PS | PM | PB |
|----|----|----|----|----|----|----|----|
| NB | NB | NB | NM | NM | NS | NS | Z  |
| NM | NB | NM | NM | NS | NS | Z  | PS |
| NS | NM | NM | NS | NS | Z  | PS | PS |
| Z  | NM | NS | NS | Z  | PS | PS | PM |
| PS | NS | NS | Z  | PS | PS | PM | PM |
| PM | NS | Z  | PS | PS | PM | PM | PB |
| PB | Z  | PS | PS | PM | PM | PB | PB |

**Table 2.** Chromosomes and Spikes (Neurons) production results

| Spikes (Neurons) I | | | |
|---|---|---|---|
| F1 | F2 | F3 | F4 |
| Spikes (Neurons) II | | | |
| F'1 | F'2 | F'3 | F'4 |
| Spikes (Neurons) III | | | |
| F''1 | F''2 | F''3 | F''4 |

The relations between *Fi, F'i, F''i* and $x_i, y_i, z_i$ are as Eq. (18).

$$\begin{pmatrix} x_1 = (F4 + F3).-0.032 \\ x_2 = F3.0.032 \\ x_3 = F1.0.032 \\ x_4 = (F_1 + F_2).0.032 \\ y_1 = (F4 + F3).-100 \\ y_2 = F3.-100 \\ y_3 = F1.100 \\ y_4 = (F1 + F2).100 \\ z_1 = (F'4 + F'3).-0.032 \\ z_2 = F'3.-0.032 \\ z_3 = F''1.0.032 \\ z_4 = (F1 + F2).0.032 \end{pmatrix} \tag{18}$$

In order to find the individual solution (optimal individual) one limits oneself to looking for the values of genes that are unknown. To reach the global optimum, large training size that is equal to 100 individuals Chromosomes and Spikes (Neurons) taken. Then GA-DSNN apply to model based on recent description of these operators and states. Stop criterion should be considered. Stop Criterion is when the maximum generation number reaches 50. The efficiency η of the algorithms was calculated using the Eq. (19).

$$\eta = \frac{\int_{t1}^{t2} P \, dt}{\int_{t1}^{t2} Pmax \, dt} \tag{19}$$

Where $t_1$ and $t_2$ are the start-up (sunrise) and system downtimes (sunset), respectively, *P* is the row efficiency power, and $P_{max}$ is the theoretical maximum reliability and low cost of hydropower system with optimized load frequency control.

## 5. Simultion Results and Discussion

At the previous part, modeling of load frequency control with Fuzzy-PD tuning with GA-DSNN proposed. This part represents the simulation results and figure. Fuzzy logic controller shown in Figure 3 and fuzzy PD input and output shown in Figure 4 and Figure 5. Moreover, 3D fuzzy surface after applying fuzzy logic to PD is shown in Figure 6.

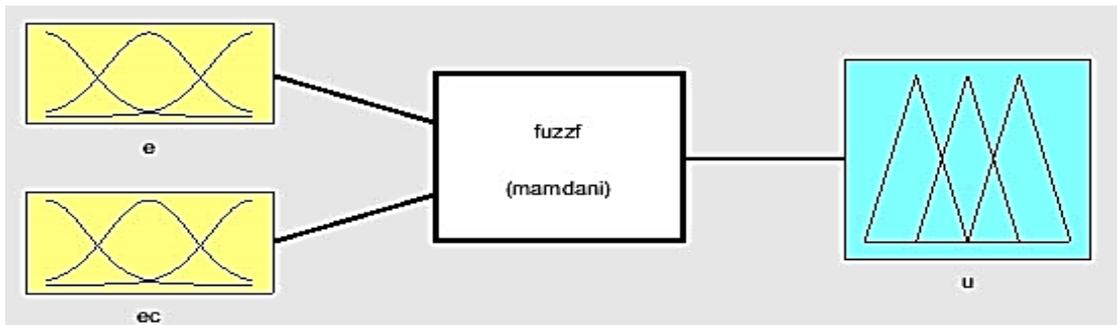

**Figure 4.** Fuzzy Logic

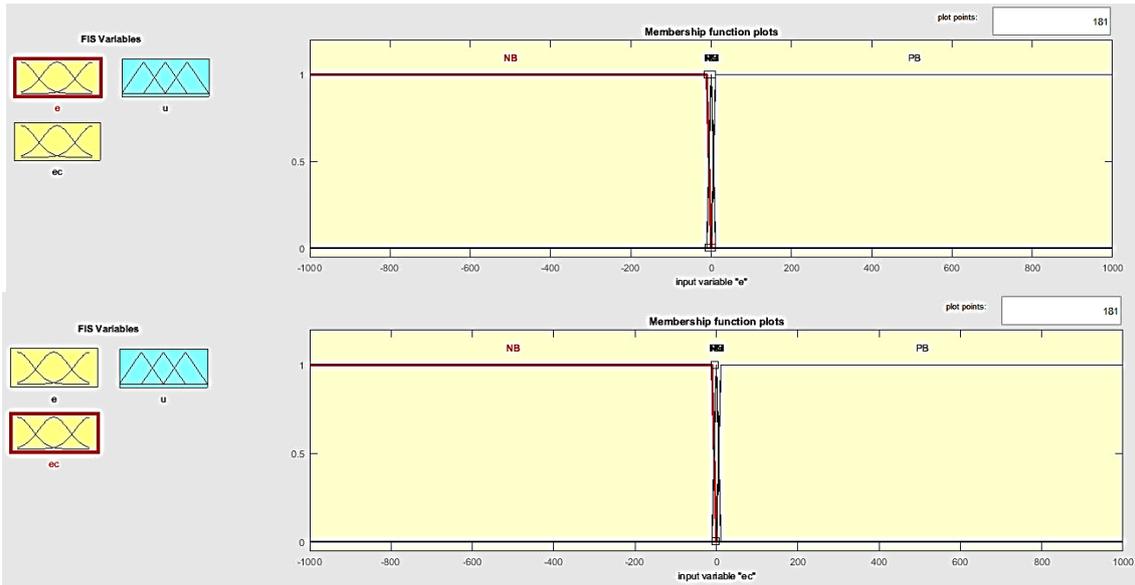

**Figure 5.** Fuzzy Inputs *(Up: e input, Bottom: ec input)*

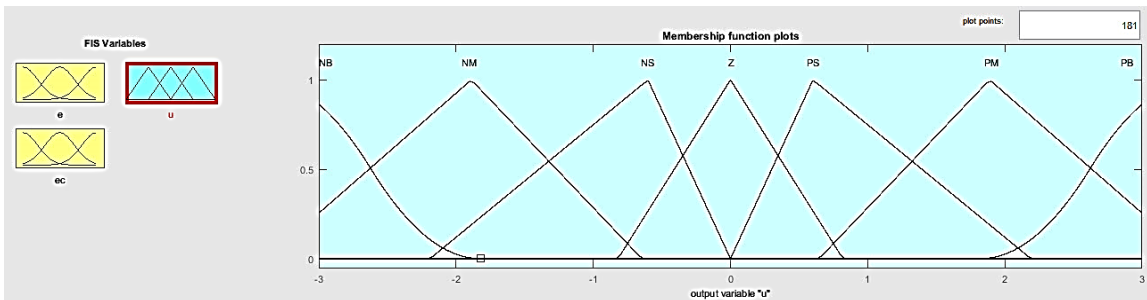

**Figure 6.** Fuzzy Output

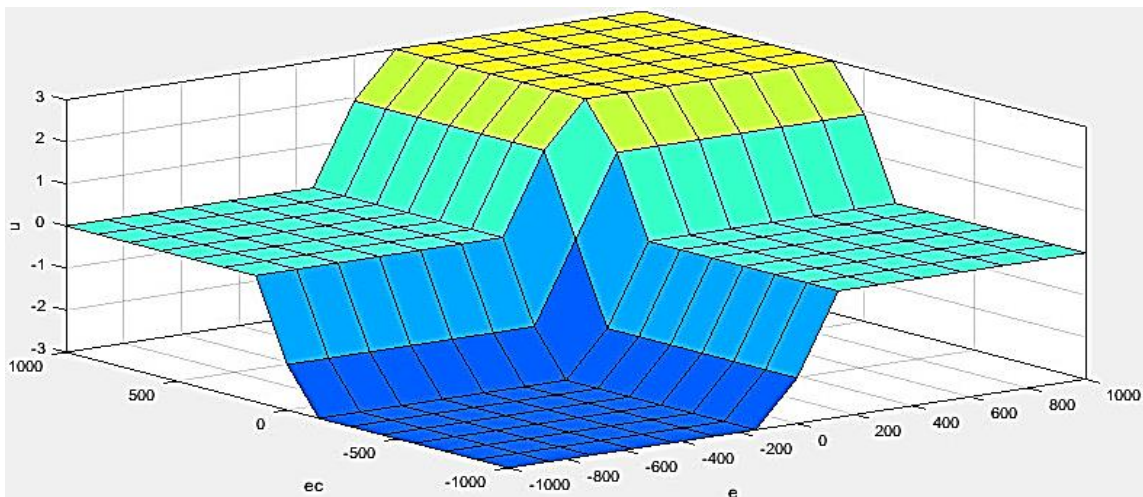

**Figure 7.** 3D Fuzzy surface after applying Fuzzy to PD controller

Table 3. Control performance measures of the load frequency control

| Performance Measures | PD Controller | | Fuzzy PD Controller | | Fuzzy-PD Tuning with GA-DSNN | |
|---|---|---|---|---|---|---|
| | 50 kW Additional Load | 50 kW Load Drop | 50 kW Additional Load | 50 kW Load Drop | 50 kW Additional Load | 50 kW Load Drop |
| Overshoot (%) | 0.554 kW | 0.756 kW | 0.1666 kW | 0.404 kW | 0.212 kW | 0.309 kW |
| Undershoot (%) | 1.333 kW | 0.484 kW | 0.409 kW | 0.150 kW | 0.281 kW | 0.117 kW |
| Setting Time $t_{0.05}$ sec | 5.132 kW | 6.316 kW | 1.299 kW | 1.545 kW | 1.001 kW | 1.184 kW |
| SSE (Hz) | 0.004 kW | 0.023 kW | 0.003 kW | 0.010 kW | 0.002 kW | 0.001 kW |
| IAE | 2.430 kW | 2.404 kW | 0.981 kW | 0.977 kW | 0.748 kW | 0.757 kW |
| ISE | 0.641 kW | 0.630 kW | 0.185 kW | 0.183 kW | 0.160 kW | 0.124 kW |
| ITAE | 9.583 kW | 9.487 kW | 3.811 kW | 3.815 kW | 2.078 kW | 2.162 kW |

Control performance measures of the load frequency control after using PD controller, fuzzy-PD and fuzzy-PD tuning with Genetic Algorithm-Deep Spiking Neural Network (GA-DSNN) is listed at Table 3. It should be noted that these performance measures of load frequency control compared between these three methods by kW nomenclature. It should be noted that GA used 100 chromosomes initially with 5 elite genes in any iteration (maximum iteration is 500 period) with the crossover rate 0.2 and mutation rate 0.02 based on random selection method.

## 6. Conclusions

In this article, a novel method proposed for load frequency control (LFC) in small scale of hydropower system based on self-tuning fuzzy-PD with GA-DSNN method. The performance of this PD controller for LFC in small scale of hydropower system is not satisfied under dynamic load conditions. So, GA-DSNN method as one of the evolutionary algorithm-deep learning models tuned and optimized the performance of fuzzy self-tuning PD controller in order to better LFC results. The proposed controller needs to compare with classic model means PD controller for LFC. In fact, PD controller, fuzzy PD controller and also fuzzy PD tuned with GA-DSNN controller compared and evaluated by each other. The obtained results represented that proposed schematic of LFC in small scale hydropower system and power plant lead to create better performance and robustness for planning estimation and prediction. The evaluation and validation methods for comparing named controller measures as performance include ITAE, ISE, IAE, SSE, overshoot and settling time.

## References

[1] H. Haggi, F. Hasanzad, and M. Golkar, "Security-Constrained Unit Commitment considering large-scale compressed air energy storage (CAES) integrated with wind power generation," *International Journal of Smart Electrical Engineering,* vol. 6, pp. 127-134, 2017.

[2] H. Shokouhandeh, M. Ghaharpour, H. G. Lamouki, Y. R. Pashakolaei, F. Rahmani, and M. H. Imani, "Optimal Estimation of Capacity and Location of Wind, Solar and Fuel Cell Sources in Distribution Systems Considering Load Changes by Lightning Search Algorithm," in *2020 IEEE Texas Power and Energy Conference (TPEC),* 2020, pp. 1-6.

[3] G. W. Frey and D. M. Linke, "Hydropower as a renewable and sustainable energy resource meeting global energy challenges in a reasonable way," *Energy policy,* vol. 30, pp. 1261-1265, 2002.

[4] F. H. Schwartz and M. Shahidehpour, "Small hydro as green power," in *2006 IEEE EIC Climate Change Conference,* 2006, pp. 1-6.

[5] O. Paish, "Small hydro power: technology and current status," *Renewable and sustainable energy reviews,* vol. 6, pp. 537-556, 2002.

[6] H. Haggi, M. Song, and W. Sun, "A review of smart grid restoration to enhance cyber-physical system resilience," in *2019 IEEE Innovative Smart Grid Technologies-Asia (ISGT Asia),* 2019, pp. 4008-4013.

[7] M. Hayerikhiyavi and A. Dimitrovski, "A Practical Assessment of the Power Grid Inertia Constant Using PMUs," *arXiv preprint arXiv:2104.07884,* 2021.

[8] L. Zhang, Q. Wu, Z. Ma, and X. Wang, "Transient vibration analysis of unit-plant structure for hydropower station in sudden load increasing process," *Mechanical Systems and Signal Processing,* vol. 120, pp. 486-504, 2019.

[9] K. P. Bharani Chandra and D. Potnuru, "A nonlinear load frequency controller for hydropower plants," *International Journal of Ambient Energy,* vol. 42, pp. 203-210, 2021.

[10] W. Yang, J. Yang, W. Zeng, R. Tang, L. Hou, A. Ma, *et al.*, "Experimental investigation of theoretical stability regions for ultra-low frequency oscillations of hydropower generating systems," *Energy,* vol. 186, p. 115816, 2019.

[11] M. R. B. Khan, J. Pasupuleti, and R. Jidin, "Load frequency control for mini-hydropower system: A new approach based on self-tuning fuzzy proportional-derivative scheme," *Sustainable Energy Technologies and Assessments,* vol. 30, pp. 253-262, 2018.

[12] X. Wang and K. Yang, "Economic load dispatch of renewable energy-based power systems with high penetration of large-scale hydropower station based on multi-agent glowworm swarm optimization," *Energy Strategy Reviews,* vol. 26, p. 100425, 2019.

[13] S. Doolla and T. Bhatti, "Load frequency control of an isolated small-hydro power plant with reduced dump load," *IEEE Transactions on Power Systems,* vol. 21, pp. 1912-1919, 2006.

[14] A. T. Hammid, M. Hojabri, M. H. Sulaiman, A. N. Abdalla, and A. A. Kadhim, "Load frequency control for hydropower plants using pid controller," *Journal of Telecommunication, Electronic and Computer Engineering (JTEC),* vol. 8, pp. 47-51, 2016.

[15] Y. Xu, C. Li, Z. Wang, N. Zhang, and B. Peng, "Load frequency control of a novel renewable energy integrated micro-grid containing pumped hydropower energy storage," *IEEE Access,* vol. 6, pp. 29067-29077, 2018.

[16] M. Mahdavian, G. Shahgholian, M. Janghorbani, B. Soltani, and N. Wattanapongsakorn, "Load frequency control in power system with hydro turbine under various conditions," in *2015 12th International Conference on Electrical Engineering/Electronics, Computer, Telecommunications and Information Technology (ECTI-CON),* 2015, pp. 1-5.

[17] R. Jamil, "Hydroelectricity consumption forecast for Pakistan using ARIMA modeling and supply-demand analysis for the year 2030," *Renewable Energy,* vol. 154, pp. 1-10, 2020.

[18] J. Chen, Y. Mei, Y. Ben, and T. Hu, "Emergy-based sustainability evaluation of two hydropower projects on the Tibetan Plateau," *Ecological Engineering,* vol. 150, p. 105838, 2020.

[19] H. Zhang, D. Chen, B. Xu, E. Patelli, and S. Tolo, "Dynamic analysis of a pumped-storage hydropower plant with random power load," *Mechanical Systems and Signal Processing,* vol. 100, pp. 524-533, 2018.

[20] A. Zargari, R. Hooshmand, and M. Ataei, "A new control system design for a small hydro-power plant based on particle swarm optimization-fuzzy sliding mode controller with Kalman estimator," *Transactions of the Institute of Measurement and Control,* vol. 34, pp. 388-400, 2012.

[21] A. Safaei, H. M. Roodsari, and H. A. Abyaneh, "Optimal load frequency control of an island small hydropower plant," in *The 3rd Conference on Thermal Power Plants*, 2011, pp. 1-6.

[22] S. Doolla and T. Bhatti, "A new load frequency control technique for an isolated small hydropower plant," *Proceedings of the Institution of Mechanical Engineers, Part A: Journal of Power and Energy,* vol. 221, pp. 51-57, 2007.

[23] X. Yu, X. Yang, C. Yu, J. Zhang, and Y. Tian, "Direct approach to optimize PID controller parameters of hydropower plants," *Renewable Energy,* vol. 173, pp. 342-350, 2021.

[24] E. J. Arbuckle, M. Binsted, E. G. Davies, D. V. Chiappori, C. Bergero, M.-S. Siddiqui*, et al.*, "Insights for Canadian electricity generation planning from an integrated assessment model: Should we be more cautious about hydropower cost overruns?," *Energy Policy,* vol. 150, p. 112138, 2021.

[25] A. Taghavirashidizadeh, R. Parsibenehkohal, M. Hayerikhiyavi, and M. Zahedi, "A Genetic algorithm for multi-objective reconfiguration of balanced and unbalanced distribution systems in fuzzy framework," *Journal of Critical Reviews,* vol. 7, pp. 639-343, 2020.

[26] T. Agarwal, F. Rahmani, I. Zaman, F. Gasbarri, K. Davami, and M. Barzegaran, "Comprehensive design analysis of a 3D printed sensor for readiness assessment applications," *COMPEL-The international journal for computation and mathematics in electrical and electronic engineering,* 2021.